# Parallel convolution processing using an integrated photonic tensor core

J. Feldmann[1,*], N. Youngblood[2,3,*], M. Karpov[4], H. Gehring[1], X. Li[2], M. Stappers[1], M. Le Gallo[5], X. Fu[4], A. Lukashchuk[4], A.S. Raja[4], J. Liu[4], C.D. Wright[6], A. Sebastian[5,#], T.J. Kippenberg[4#], W.H.P. Pernice[1,7,#] and H. Bhaskaran[2,#]

[1] Institute of Physics, University of Muenster, Heisenbergstr. 11, 48149 Muenster, Germany
[2] Department of Materials, University of Oxford, Parks Road, OX1 3PH Oxford, UK
[3] Current Address: Department of Electrical and Computer Engineering, University of Pittsburgh, 3700 O'Hara St., Pittsburgh, PA 15261, USA
[4] Laboratory of Photonics and Quantum Measurements, Swiss Federal Institute of Technology Lausanne (EPFL), Station 3, CH-1015, Lausanne, Switzerland
[5] IBM Research - Zurich, Säumerstrasse 4, 8803 Rüschlikon, Switzerland
[6] Department of Engineering, University of Exeter, Exeter, EX4 4QF, UK
[7] Center for Soft Nanoscience, University of Münster, 48149 Münster, Germany

\* These authors contributed equally.

#Correspondence to: wolfram.pernice@uni-muenster.de, harish.bhaskaran@materials.ox.ac.uk, kippenberg@epfl.ch, ASE@zurich.ibm.com

**With the proliferation of ultra-high-speed mobile networks and internet-connected devices, along with the rise of artificial intelligence, the world is generating exponentially increasing amounts of data—data that needs to be processed in a fast, efficient and 'smart' way. These developments are pushing the limits of existing computing paradigms, and highly parallelized, fast and scalable hardware concepts are becoming progressively more important. Here, we demonstrate a computational specific integrated photonic tensor core—the optical analog of an ASIC—capable of operating at Tera-Multiply-Accumulate per second (TMAC/s) speeds. The photonic core achieves parallelized photonic in-memory computing using phase-change memory arrays and photonic chip-based optical frequency combs (soliton microcombs). The computation is reduced to measuring the optical transmission of reconfigurable and non-resonant passive components and can operate at a bandwidth exceeding 14 GHz, limited only by the speed of the modulators**

**and photodetectors. Given recent advances in hybrid integration of soliton microcombs at microwave line rates, ultra-low loss silicon nitride waveguides, and high speed on-chip detectors and modulators, our approach provides a path towards full CMOS wafer-scale integration of the photonic tensor core. While we focus on convolution processing, more generally our results indicate the major potential of integrated photonics for parallel, fast, and efficient computational hardware in demanding AI applications such as autonomous driving, live video processing, and next generation cloud computing services.**

The increased demand for machine learning on very large datasets[1] and the growing offering of artificial intelligence services on the cloud[2–4] has driven a resurgence in custom hardware designed to accelerate multiply and accumulate (MAC) computations—the fundamental mathematical element needed for matrix-vector multiplication (MVM) operations. Whilst various custom silicon computing hardware (i.e. FPGAs[5], ASICs[6], and GPUs[7]) have been developed to improve computational throughput and efficiency, they still depend on the same underlying electronic components, which are fundamentally limited in both speed and energy by Joule heating, RF crosstalk, and capacitance[8]. The last of these (capacitance) dominates energy consumption and limits the maximum operating speed in neural network hardware accelerators[9]. This is because, the movement of data (e.g. trained network weights), rather than arithmetic operations, requires the charging and discharging of chip-level metal interconnects. Thus, improving the efficiency of logic gates at the device level provides diminutive returns in such applications, if the flow of data during computation is not simultaneously addressed[10]. Even recent developments such as memristive crossbar arrays[11–14] to compute in the analog domain, whilst promising, do not have the potential for parallelizing the MVM operations (save for physically replicating the elements of the matrix). Moreover, they are plagued by the same limitations of electronic addressing[15], with additional challenges in the manufacturing and implementation due to issues with device variability[16,17], cyclability[18], and drift[19,20].

Integrated photonics benefits from the modularity and scalable fabrication methods of integrated circuits, whilst having two key advantages over its electronic counterparts: (1) massively parallel data transfer through wavelength division multiplexing (WDM) in conjunction with multichannel sources (i.e. optical frequency combs); and (2) extremely high data modulation speeds limited only by the bandwidth of on-chip optical modulators and photodetectors. These uniquely photonic advantages have led to the ubiquity of optical networks for information transfer and are presently revolutionizing data centre interconnects (i.e. server-to-switch communication). However, these developments have yet to seriously challenge digital electronics in the arena of information processing. Despite the current dominance of integrated electronics for computing, an application-specific optical processor not limited by the energy-bandwidth trade-off of electrical interconnects[8] could bring the advantages of optical networking to the field of computing. This would result in very high computational throughput via low-latency (i.e. information processing and propagation at the speed of light) and parallel operations in a single physical optical processing core using WDM — essentially providing an additional scaling dimension through use of frequency space. While the concept of free-space optics for efficient linear computing (e.g. Fourier transforms, convolutions, matrix multiplication, etc.) has existed for many decades[21] and continues to inspire novel computing architectures[22–26], precise control of the optical phase over the entire system remains the primary factor limiting scalability and commercialization.

Integrated photonics holds promise to solve these challenges. However, integration together with CMOS compatible manufacturing is of paramount importance: on chip, both energy-efficient optical memory units and a compact, broadband multi-channel laser source must be combined within a scalable photonic architecture. Recent work on integrated photonic processors for MVMs and neuromorphic computing[27–29] has shown the potential advantages of the photonic approach, but key issues such as large footprints (11,000 $\mu m^2$ per interferometer unit[27]) and the use of thermo-optic heaters to tune the phase or resonance wavelength of their components (ranging on average from 1 mW to 10 mW per heater for ring resonators and Mach-Zehnder interferometers respectively) were bottlenecks[30], as well as resonant devices such as add-drop resonators that limit the modulation bandwidth. Additionally, while using WDM for processing multiple inputs simultaneously in the same physical hardware has been proposed[31], it has not yet been demonstrated on-chip.

Here, we design and experimentally demonstrate a novel scalable, CMOS compatible, photonic hardware accelerator (which we term "photonic tensor core" in the following) capable of many parallel MVM operations at optical data rates to process images using convolution filters (here, edge detection and emboss filters) and test it on the MNIST database with a small-scale convolutional neural network (CNN). In a departure from electronic accelerators (see Fig. 1a), our photonic processor implements an on-chip matrix multiplication engine capable of performing parallel multiply-accumulate operations using multiple wavelengths derived from a photonic chip-based optical frequency comb, that are incoherently added within a network of waveguides that exploit phase-change materials. We leverage recent advances in chip-scale microcombs[32,33] operating in the regime of dissipative Kerr soliton (DKS) states, which enable broadband, low-noise, and fully integrated optical frequency combs with line spacing ranging from GHz to THz domains and that are compatible with wafer scale manufacturing and integration with on-chip lasers[34–36]. These devices have already been employed in system level demonstrations such as massively parallel coherent communications[37], chip-scale frequency synthesizers[38], and massively parallel LiDAR[39]. Thus far DKS systems, however, have remained unexplored for photonic computing.

Key to our approach is the encoding of image data onto the individual comb teeth of an on-chip frequency comb, and subsequently encoding fixed convolutional kernels in the non-volatile configuration (i.e. the amorphous or crystalline phase) of integrated phase-change material cells that couple evanescently to a matrix of interconnected photonic waveguides

(shown in Fig. 1b). Our approach minimizes both latency and the movement of data, by using non-volatile in-memory photonic MAC operations and greatly reduces the footprint cost of photonics by multiplexing computations in the same photonic core. Importantly, both the soliton microcombs and the matrix of photonic waveguides can be implemented in silicon nitride[40], an ultra-low loss, CMOS compatible nonlinear integrated photonic platform, that is compatible with wafer scale manufacturing and foundry. Combined with recent advances in both on chip modulators and hybrid integration of soliton microcombs[34,35], fully integrated custom photonic tensor cores are viable.

**Realization of parallel 2D convolutions via matrix-vector multiply operations**

One prominent class of machine learning models that stand much to gain from high throughput accelerators are convolutional neural networks (CNNs) which are highly effective for applications such as in image classification, autonomous navigation, and audio analysis in the frequency domain. In state-of-the-art CNNs, many convolution "hidden layers" are applied to an input signal before feeding the processed data to fully connected layers for classification[41,42]. Each of the convolution layers takes in an input image, performs convolutional operations to extract features, and generates an output image. For the case of a convolution between an input image of dimension $n \times n$ with $d_{in}$ channels and a filter of dimension $k \times k \times d_{in}$, the resulting output image is of dimension $(n-k+1) \times (n-k+1)$. To perform each convolution operation, a filter is passed over the input image inspecting a small window of pixels at a time. A pixel-wise MAC operation between the filter and the current filter window is carried out to calculate a single pixel of the output image. In CNNs, $d_{out}$ convolution kernels will be applied to the same image, which corresponds to $(n-k+1)^2 \times k^2 \times d_{in} \times d_{out}$ MAC operations per convolution layer and scales in computational complexity as $O(n^2 k^2)$. It is worth noting that for the case of large kernels ($k > 15$), performing the convolution in the Fourier domain can reduce computational complexity[43] to $\sim O(n^2 \ln(n))$. However, $k \leq 5$ for most kernels in many common CNN models used today

(i.e. AlexNet[44], ResNet[41], GoogLeNet[45], etc.) making the Fourier approach less efficient than direct convolution. When performing convolution operations in the digital domain, a minimum of two clock cycles are required for each sequential MAC operation—though the number of clock cycles for floating point multiplication is usually $\geq 3$[46]. This leads to a significant computational bottleneck, requiring distribution across multiple computing cores as illustrated in Fig. 1a.

In order to build efficient hardware to perform the convolution operations, one approach (originally conceived for electronic in-memory computing using memristive crossbar arrays[47,48]) is to combine all the convolutional filters into a large filter matrix stored in memory. As depicted in Fig. 1c, the filter matrix will be of dimension $(k^2 \times d_{in}) \times d_{out}$. It is constructed by stacking the kernel matrices into the columns of the final filter matrix. In the same way the pixels of the input image are rearranged by stacking the pixels of the filter volume, $(k \times k \times d_{in})$, into the rows of the input matrix. Hence a single convolution operation involves $(n-k+1)^2$ MVM operations between the filter matrix and the input vectors of $k^2 \times d_{in}$ dimension. In the electronic domain, these MVM operations are typically multiplexed in time (serial processing) with parallelization afforded only by physically replicating the filter matrix. In this work, we exploit photonic integrated soliton microcomb and optical WDM to overcome this fundamental limitation by encoding multiple input vectors of dimension $k^2 \times d_{in}$ onto multiple lines of a coherent chip-scale frequency comb. These optical input vectors can then be applied to a single $(k^2 \times d_{in}) \times d_{out}$ filter matrix simultaneously, thus eliminating duplicated physical hardware and sequential operations. This approach will be employed when designing the photonic tensor core.

**The photonic tensor core**

First, we demonstrate how to perform an MVM operation in the optical domain using photonic integrated circuits employing non-volatile phase-change cells that store analog values of the matrix *in situ*[49]. Details using phase-change materials (PCMs) on single devices are described elsewhere [49,50]. In this work, the PCM ($Ge_2Sb_2Te_5$ or GST) cells are employed as attenuating matrix elements which absorb a desired amount of light depending on their particular phase configuration. In the crystalline PCM state, most of the incoming light is absorbed, representing for example a "0". In the amorphous state, most of the light is transmitted, thus representing a "1". Intermediate transmission states can be chosen by controllably switching fractions of amorphous and crystalline parts in the PCM cell[49,51]. To achieve both positive and negative matrix elements, we define "0" as an intermediate state between the crystalline and amorphous states as described in the Supplementary Information.

In order to calculate the $m \times n$ MVM operation shown at the top of Fig. 2a, the input vector is encoded in the amplitude of the optical signals sent to the different matrix inputs. In addition to amplitude at a given wavelength, the input vector is also encoded at different wavelengths providing the ability for multiple calculations to be carried out simultaneously, while avoiding unwanted interference at the photodetector array. The amplitude of each wavelength represents one of the vector entries ($X_1, ... X_m$). Therefore, the input vectors can be fed to the matrix by modulating the input signals with currently available fast electro-optical modulators, providing access to very high data rates. The matrix itself is designed as a waveguide crossbar array with additional directional couplers that equally distribute the input power to all PCM-cells (more details of the splitting ratios of the directional couplers are given in the supplementary information). By using a soliton microcomb with a mode spacing that exceeds the detector bandwidth, interference inside the waveguides can be avoided and the summation of the individual products (of the matrix-vector multiplications) can be performed by adding the comb teeth to the output waveguides, also by using directional couplers. With the horizontal directional couplers, the input vectors are equally distributed to the different columns of the matrix (which represent the individual image kernels) whereas the vertical directional couplers combine the input light after interaction with the PCM cells and perform the accumulation operation. It should be noted that each vector entry only interacts with a single PCM cell per matrix column. This interaction can be viewed as a single multiplication between the incoming amplitude and the absorption of the phase-change cell, as has been shown in previous work[52]. The output power at each column of the matrix represents the inner-product (the sum of the individual products) of the input vector with a kernel multiplied by a certain (fixed) factor $\left(\frac{1}{mn}\right)$ which depends on the matrix size. Power distribution due to fan-out accounts for the $1/n$ loss, while combining $m$ non-interfering sources with directional couplers accounts for the additional $1/m$ loss due to energy conservation.

Figure 2b depicts a scanning electron micrograph of the resonator used for comb generation and Fig. 2c shows an optical image of a fabricated 16x16 matrix with a 4x4 matrix as an inset. Key chip regions are magnified in the scanning-electron micrographs on the right. Coupling of light into the optical chip is achieved using broadband total internal reflection (TIR) couplers[53,54] (bottom right of Fig. 2c). The TIR couplers provide access to a wide wavelength spectrum and thus allow the coupling of multiple wavelengths into the chip (more details are given in Supplementary Information, section 12). The PCM-cells (of area 3×3 µm²) acting as the matrix elements are deposited on top of waveguide crossings (Fig. 2c top right). Each individual matrix cell has three additional grating couplers used to optically address the PCM. By sending pulses (via the middle coupler) to the waveguide directly leading to the PCM cell on the crossing, it can be optically switched for programming each matrix element (in this case the light is coupled to the chip using Bragg-grating couplers because operation at a single wavelength (1550 nm) is sufficient).

In addition to substantial benefits in modulation speed (for changing the vector inputs), an optical implementation of a matrix-vector multiplier allows the harnessing of wavelength division multiplexing (MUX) to execute parallel MVM operations. In particular, as Fig. 2d shows, the same matrix can be used to process several input vectors at the same time when all the individual vectors are encoded in different wavelengths. For the 4×4 matrix example shown in Fig. 2, and the processing of four input vectors per time step, sixteen different wavelengths are needed. In this work, these wavelengths are generated using a single DKS state of a microcomb[33,55,56] which is fed into a demultiplexer to split up the individual wavelengths ($\lambda_1$–$\lambda_{16}$). After manipulating the amplitude of each comb line individually (according to the value of the input vectors) by using variable optical attenuators (VOAs), the corresponding entries of each vector are multiplexed back together (i.e. $\lambda_1$, $\lambda_5$, $\lambda_9$, $\lambda_{13}$) and sent to the matrix input. After propagating through the filter matrix, all output waveguides of the matrix contain all 16 input wavelengths. Proper demultiplexing and combining of the wavelengths corresponding to the individual vectors yields the convolution results that can be measured with photodetectors. In the current example, 16 inner-product operations (four kernels applied to four input vectors) are carried out in a single time step. Depending on the number of lines available in the frequency comb, the multiplexing scheme can be extended further leading to significant speed gains. Figure 2e shows the optical spectrum of an on-chip microcomb revealing lines with 100 GHz spacing over a range of more than 25 Thz.

To illustrate the principle outlined above experimentally, the convolution of an input image depicting a handwritten "4" (Fig. 3a)[57] is performed using four 3×3 image kernels (resulting in a 9×4 filter matrix) and a single vector (9×1) per time step (Fig. 3b-e). Note that, $d_{in} = 1$ and $d_{out} = 4$ in this example. The image kernels applied in this example are chosen for edge detection and are shown below the output images (for more details on how the matrix elements are exactly defined in the PCM-state, see Supplementary Information, section 8). After obtaining the results of the matrix-vector multiplications, the output values are offset by +0.5 and the values below 0 are set to 0 (black pixel) and the values above 1 are set to 1 (white).

Each of the kernels highlights different edges of the original image: Fig. 3b, for example, highlights upper edges, whereas Fig. 3d brings out the opposite lower edges. Fig. 3f shows the combined images (difference between alternating edges and addition of the two resulting images) highlighting that all edges have been properly detected. Since the four kernels are all inscribed in the same matrix, the pixel values of all four output images are obtained simultaneously, including more than 63,000 inner-product operations in total. The edge features are strongly visible, which emphasizes the effectiveness of our optical convolution operation. The variation in the background is due to power fluctuations of the comb lines over time, leading to small errors in the matrix-vector multiplications. The inner-product of the entire convolution was processed at approximately 1 kHz limited only by the speed of the VOAs. Thus, due to the slow electronic control and serial communication between the computer and microcontroller, the overall processing in this particular example took about four minutes. It should be noted that in the examples of Fig. 3b-e, for each optical matrix-vector multiplication, an MVM operation in software is also performed in a post-processing step to subtract a certain reference power from the measured output power in the matrix columns (more details are provided in the supplementary information).

In order to avoid this post-processing, the reference convolution operation can also be performed optically in the same on-chip matrix. In this case, one matrix column is in a reference state (for example all PCM-cells in the crystalline phase state). The output value from this column is then subtracted from all the matrix columns holding the actual image kernels. Figures 3g-i show an experimental example of a convolution operation which was performed without electrical post-processing using reference subtraction. Here, a 3×3 kernel (emboss filter) was applied using a 9×2 matrix, with one column for the image kernel and one column for the reference. The original image is shown on the left, while the experimental output image after the convolution operation is shown in the middle panel. From comparison with the calculated expected output on the right, it can be seen that the on-chip matrix also performs well without the need for the post-processing step. It should be noted, that even though the image has three color channels red, green and blue ($d_{in} = 3$), the convolutions are performed on each channel independently and combined in the end leading to the output image – however, this is more a limitation of the size of our hardware matrix as opposed to a fundamental limitation of this technology.

Having demonstrated the basic capabilities of our phase-change integrated photonic approach to performing convolution operations in the optical domain, we now show, in Fig. 4, experimental examples of processing four input vectors in parallel at the same time. In this case, four pixels of the new image are obtained per image kernel simultaneously, therefore shortening the processing time by a factor of four. The kernel size used for this experiment is 2×2 and the input dimension of the image, $d_{in} = 1$, leading to a 4×4 filter matrix. The convolutions again highlight different edges which can be clearly seen—for example, in the representation of the bricks in the upper image. Panel b) and e) emphasize vertical edges, whereas panels c) and d) highlight horizontal edges. This is in spite of variations in the vertical direction due to power fluctuations of the input signal, underlining the robustness of the technique. Panel f) shows the combined images highlighting all edges. As four vectors are processed in parallel, also the processing time is decreased here from four to approximately one minute.

**Digit recognition with a convolutional neural network**

Having shown that the photonic tensor core is capable of processing the convolutions demonstrated with different image filters, in a next step a convolutional neural network is built and tested against the MNIST handwritten digit database. The CNN employed in our experiments is depicted in Fig. 5a and consists of the input layer taking the pixel data (28x28 pixels, single channel) that is then passed to a convolution layer consisting of four 2x2 kernels plus subsequent Rectified Linear Unit (ReLU) activation, resulting in an output of dimension 27x27x4 (valid padding). The output from the convolution step is flattened and fed to a fully-connected layer with ten neurons. The probabilities for every digit are obtained from the final classification using the softmax function. The network was trained via software (see section 14 of Supplementary Information for more detail) and the weights of the filter kernels programmed to the states of the PCM-cells in the on-chip matrix. To test the accuracy of the predictions of the network, 10,000 test images were processed using the photonic matrix for the convolutions at a rate of 2 GHz (resulting in processing time of 8.1 µs per image) with an FPGA for electronic control (more details on the experimental setup are given in the Supplementary Information). The confusion matrices illustrating the predictions for the different images with the experimentally obtained and the calculated results are shown in Fig. 5b. The experimental implementation of the CNN reached an accuracy of 95.3% showing good agreement with the calculated prediction accuracy of 96.1%.

To analyze the computational accuracy of the optical convolution processor for single dot-product operations, randomly chosen input vectors with nine entries are processed using a fixed matrix column and are compared against the expected analytically calculated multiplication result. The results for 100,000 calculations are scaled to the range [0,1] and plotted in Fig. 5c together with the corresponding histogram revealing a standard deviation of 0.008 resulting in a resolution of 5 bits (more information on the evaluation of the resolution is given in the supplementary information). This experiment was carried out using the same setup as for edge detection using VOAs operated in the kHz regime and a 9x1 matrix. We note, however, that the main source limiting the precision of matrix vector multiplications in our architecture are the electronic signals driving the modulators and the extinction ratio of the modulator (which both do not depend on frequency). Therefore, no loss in precision is expected when driving the system at higher speeds. We also note, that the matrix elements can be programmed with a precision of more than 8-bit using a closed-loop approach, as shown in the supplementary figure S6.

**Projections to the future**

The above data was obtained with matrices up to a size of 9×4, with maximal four input vectors per time step and a modulation speed of up to 2 GHz. To estimate the ultimate performance capabilities of the system, we now explore the scaling capabilities in terms of matrix size, modulation speed and number of parallel vectors. The main factor impacting the achievable matrix size is the optical loss induced by the photonic matrix, which results from equally splitting the light to all matrix cells, combined with the insertion loss of the directional couplers and waveguide crossings. As detailed in the description of the construction of the photonic tensor core (detailed in Supporting Information), the optical loss in the matrix itself scales with the matrix size as 1/(m×n) for a matrix size m×n. Additional loss is added by the directional couplers and the waveguide crossings and increases linearly with the matrix size. The propagation loss of the waveguides (0.2 dB/cm) can be neglected in comparison to these contributions. Figure 5d) shows a heatmap of the calculated matrix loss as a function of the matrix size, considering measured insertion loss of 0.1 dB per directional coupler and 0.12 dB per crossing (see Supplementary Information, sections 4 and 6). The stars on the diagonal are measured optical loss for fabricated matrices with sizes up to 32×32, and agree well with the calculations. By further improving the loss of the crossings[58] and directional couplers the limits to matrix sizes can be increased.

To illustrate convolution processing using high-speed modulation of the input vectors, Fig. 5e shows an eye diagram at 13.5 GHz modulation speed obtained from a 2x1 matrix. The two electro-optical input modulators (40 GHz bandwidth) were driven by $2^7$ Pseudo-random-bit-patterns (PRBS) provided by a fast pulse-generator, therefore resulting in three output levels that can be clearly distinguished. As the photonic matrix itself is operated passively in a transmission measurement, the speed is only limited by the bandwidth of the modulators and detectors. In the experiment a detector with a -3 dB bandwidth of 12 GHz was used (additional data of modulating the individual matrix inputs up to 14 GHz is included in the supplementary material).

Because the photonic system is designed with broadband input couplers and broadband directional couplers in silicon nitride with a wide optical transparency window, the tensor processor supports more than 200 individual wavelengths from the frequency comb source with a spacing of 100 GHz (see Supplementary Information, section 13). In addition to the spectral width of the frequency comb, the influence of wavelength dependent parts in the matrix design must also be considered when estimating the wavelength range exploitable for the calculations. In this case, it is predominantly the wavelength dependence of the directional couplers that hinders the equal distribution of the input power for all wavelengths. While our design offers an impressive range of approximately 100 nm, this can be considerably improved by an adapted design[59]. The influence of dispersion in the PCM absorption can be neglected in the considered wavelength range and could be corrected by adjusting the input amplitudes of the different comb lines. Thus, for a 9×4 matrix, four multiplexed input vectors and a modulation speed of 14 GHz, a processing speed of 2 TMAC/s (9×4 MACs × 4 input vectors × 14 GHz) can be obtained. This, however, is not the ultimate speed, since we are limited here by the modulation and detection bandwidth of our particular experimental setup.

When comparing optical architectures with digital electronics, it is helpful to use compute density (defined here as TOPs (tera-operations) per second normalized by the processor area[60]) as a figure-of-merit for performance. This helps to directly compare the processing throughput of architectures that may employ very different schemes for computing MVM operations. For the SiN devices demonstrated here, the area of a single MAC (with one MAC being two operations) unit cell is 285 µm × 354 µm. This, when operating at 12 GHz with 4 input vectors via WDM, corresponds to a compute density of 1.2 TOPS/mm$^2$. By moving to a silicon-on-insulator platform with a nominal bend radius of 5 µm and using integrated electrical control of the GST[61,62], it would be straight-forward to reduce the area of the MAC unit cell to less than 30 × 30 µm$^2$, increasing the compute density to 420 TOPS/mm$^2$ per input channel (see Supplementary Information, section 11. We also demonstrate increased compute density with an SOI prototype illustrating the feasibility of this approach.) and scaling linearly with the number of input vectors via WDM—a notably different computing paradigm compared to electronic approaches (note that the compute density considers only the photonic tensor core itself, without the electronic control and the off-chip multiplexers). The energy efficiency for the actual experiments can be calculated to be 0.4 TOPS/W for 5-bit resolution (including the optical power as well as the ADCs and modulators which we estimate to be dominating the consumption, see supplementary materials). Moreover, by reducing the loss of the directional couplers and waveguide crossings and integrating detectors and modulators on-chip, the efficiency can be increased to 7.0 TOPS/W in the future. Considering only the optical energy based on the power needed to overcome shot-noise for a fixed 8-bit precision number at the output, the energy per MAC operation can be as low as 17 fJ/MAC.

To estimate the full capabilities of the optical accelerator for convolution operations, the performance of common optical components in foundry services[63,64] must be considered in combination with the wavelength range of the frequency comb that can be used. The frequency comb clearly shows lines from 1500 nm to 1650 nm (see supplementary information), leading to a range of 150 nm that could be exploited for computation that can even be extended by optimizing the setup. Considering the spacing of the comb lines of 100 GHz (0.8 nm), this leads to approximately 150 nm / 0.8 nm = 187 different wavelengths. Decreasing the spacing to 50 GHz (0.4 nm) and increasing the matrix size to 50×50, the operational speed can reach an unprecedented 1 PMAC/s for a single matrix , assuming a modulation and detection speed of 50 GHz. These large matrix sizes are experimentally feasible using variable-length directional couplers and have been demonstrated using a photonics foundry process in 2013 by J. Sun et al[65].

**Conclusion**

We describe the first instance of a photonic tensor core which combines in-memory computing with state-of-the-art photonic integrated frequency combs enabling parallelizing convolution operations in the same physical device. We demonstrate the simultaneous data transfer and computing at speeds comparable to fiber networks. Prior optical approaches to computing have largely been limited by a lack of integrated non-volatile photonic memory and the lack of multiplexing capability for such calculations[27,29,60]. Our approach overcomes both these limitations by using nonvolatile, phase-change materials integrated on waveguides to locally store convolution kernels on-chip, and photonic chip-based frequency combs which enables true in-memory photonic computing using WDM capability. The photonic tensor core demonstrated in this work is capable of operating at the speed of 2 TMAC/s, promising even faster operation by an increase of several orders of magnitude by moderate scaling with state-of-the-art foundry processes. A key feature of our approach is that, because the convolution operation is a passive transmission measurement, the calculations can in theory be performed at the speed of light at very low power, experimentally limited only by the modulation and detection bandwidths. Making use of the wavelength division multiplexing capabilities inherent to all-optical systems, our fast and parallelized implementation promises higher computational bandwidths when compared to electronic devices, as several pixels or even complete images can potentially be processed in a single time step. Our approach for convolution processing provides an effective method to remove the computing bottleneck in machine learning hardware for applications ranging from live video processing to autonomous driving and AI-aided life-saving applications. More importantly, such an approach more broadly suggests that integrated photonics are coming of age and in some cases can begin to match and even challenge electronic computation.

# References

1. Ben-Nun, T. & Hoefler, T. Demystifying parallel and distributed deep learning: An in-depth concurrency analysis. *ACM Comput. Surv.* **52**, (2019).

2. Amazon AWS Machine Learning. https://aws.amazon.com/machine-learning/.

3. Google Cloud. https://cloud.google.com/products/machine-learning/.

4. Microsoft Azure. https://azure.microsoft.com/en-us/overview/machine-learning/.

5. Zhang, C. *et al.* Optimizing FPGA-based Accelerator Design for Deep Convolutional Neural Networks. *ACM/SIGDA Int. Symp. Field-Programmable Gate Arrays(FPGA)* 161–170 (2015) doi:10.1145/2684746.2689060.

6. Jouppi, N. P. *et al.* In-Datacenter Performance Analysis of a Tensor Processing Unit. *Proc. ISCA '17* (2017) doi:10.1145/3079856.3080212.

7. Wang, P. S., Liu, Y., Guo, Y. X., Sun, C. Y. & Tong, X. O-CNN: Octree-based convolutional neural networks for 3D shape analysis. *ACM Trans. Graph.* **36**, (2017).

8. Miller, D. A. B. Attojoule Optoelectronics for Low-Energy Information Processing and Communications. *J. Light. Technol.* **35**, 346–396 (2017).

9. Agrawal, S. R. *et al.* A Many-core architecture for in-memory data processing. *Proc. Annu. Int. Symp. Microarchitecture, MICRO* 245–258 (2017) doi:10.1145/3123939.3123985.

10. Miller, D. A. B. Are optical transistors the logical next step? *Nat. Photonics* **4**, 3–5 (2010).

11. Ielmini, D. & Wong, H. S. P. In-memory computing with resistive switching devices. *Nat. Electron.* **1**, 333–343 (2018).

12. Le Gallo, M. *et al.* Mixed-precision in-memory computing. *Nat. Electron.* **1**, 246–253 (2018).

13. Boybat, I. *et al.* Neuromorphic computing with multi-memristive synapses. *Nat.*

*Commun.* **9**, (2018).

14. Sebastian, A., Le Gallo, M., Khaddam-Aljameh, R. & Eleftheriou, E. Memory devices and applications for in-memory computing. *Nat. Nanotechnol.* **15**, 529–544 (2020).

15. Hu, M. *et al.* Dot-product engine for neuromorphic computing: Programming 1T1M crossbar to accelerate matrix-vector multiplication. *Proc. - Des. Autom. Conf.* (2016) doi:10.1145/2897937.2898010.

16. Gong, N. *et al.* Signal and noise extraction from analog memory elements for neuromorphic computing. *Nat. Commun.* **9**, (2018).

17. Joshi, V. *et al.* Accurate deep neural network inference using computational phase-change memory. *Nat. Commun.* **11**, 1–13 (2020).

18. Yang, T. Y., Park, I. M., Kim, B. J. & Joo, Y. C. Atomic migration in molten and crystalline Ge2 Sb2 Te5 under high electric field. *Appl. Phys. Lett.* **95**, (2009).

19. Koelmans, W. W. *et al.* Projected phase-change memory devices. *Nat. Commun.* **6**, (2015).

20. Kim, S. *et al.* A phase change memory cell with metallic surfactant layer as a resistance drift stabilizer. *2013 IEEE Int. Electron Devices Meet.* 762–765 (2013) doi:10.1109/IEDM.2013.6724727.

21. Bell, T. E. Optical computing: A field in flux: A worldwide race is on to develop machines that compute with photons instead of electrons � but what is the best approach? *IEEE Spectr.* **23**, 34–38 (1986).

22. Hamerly, R., Sludds, A., Bernstein, L., Soljačić, M. & Englund, D. Large-Scale Optical Neural Networks based on Photoelectric Multiplication. **021032**, 1–12 (2018).

23. Dong, J., Rafayelyan, M., Krzakala, F. & Gigan, S. Optical Reservoir Computing Using Multiple Light Scattering for Chaotic Systems Prediction. *IEEE J. Sel. Top. Quantum Electron.* **26**, 1–12 (2019).

24. Silva, A. *et al.* Performing mathematical operations with metamaterials. *Science (80-. ).*

**343**, 160–163 (2014).

25. Lin, X. *et al.* All-optical machine learning using diffractive deep neural networks. *Science (80-. ).* **361**, 1004–1008 (2018).

26. Colburn, S., Chu, Y., Shilzerman, E. & Majumdar, A. Optical frontend for a convolutional neural network. *Appl. Opt.* **58**, 3179–3186 (2019).

27. Shen, Y. *et al.* Deep learning with coherent nanophotonic circuits. *Nat. Photonics* **11**, 441–446 (2017).

28. Tait, A. N. *et al.* Silicon Photonic Modulator Neuron. *Phys. Rev. Appl.* **11**, (2019).

29. Pérez, D. *et al.* Multipurpose silicon photonics signal processor core. *Nat. Commun.* **8**, (2017).

30. Galal, S. & Horowitz, M. Energy-efficient floating-point unit design. *IEEE Trans. Comput.* **60**, 913–922 (2011).

31. Bangari, V. *et al.* Digital Electronics and Analog Photonics for Convolutional Neural Networks (DEAP-CNNs). *IEEE J. Sel. Top. Quantum Electron.* **26**, (2020).

32. Herr, T. *et al.* Temporal solitons in optical microresonators. *Nat. Photonics* (2014) doi:10.1038/nphoton.2013.343.

33. Herr, T., Gorodetsky, M. L. & Kippenberg, T. J. Dissipative Kerr Solitons in Optical Microresonators. *Nonlinear Opt. Cavity Dyn. From Microresonators to Fiber Lasers* **8083**, 129–162 (2015).

34. Raja, A. S. *et al.* Electrically pumped photonic integrated soliton microcomb. *Nat. Commun.* (2019) doi:10.1038/s41467-019-08498-2.

35. Stern, B., Ji, X., Okawachi, Y., Gaeta, A. L. & Lipson, M. Battery-operated integrated frequency comb generator. *Nature* (2018) doi:10.1038/s41586-018-0598-9.

36. Jones, R. *et al.* Heterogeneously Integrated InP/Silicon Photonics: Fabricating fully functional transceivers. *IEEE Nanotechnol. Mag.* (2019) doi:10.1109/MNANO.2019.2891369.

37. Marin-Palomo, P. *et al.* Microresonator-based solitons for massively parallel coherent optical communications. *Nature* **546**, 274–279 (2017).

38. Spencer, D. T. *et al.* An optical-frequency synthesizer using integrated photonics. *Nature* **557**, 81–85 (2018).

39. Riemensberger, J. *et al.* Massively parallel coherent laser ranging using soliton microcombs. 1–18 (2019).

40. Moss, D. J., Morandotti, R., Gaeta, A. L. & Lipson, M. New CMOS-compatible platforms based on silicon nitride and Hydex for nonlinear optics. *Nature Photonics* (2013) doi:10.1038/nphoton.2013.183.

41. He, K., Zhang, X., Ren, S. & Sun, J. Deep residual learning for image recognition. *Proc. IEEE Comput. Soc. Conf. Comput. Vis. Pattern Recognit.* 770–778 (2016) doi:10.1109/CVPR.2016.90.

42. Simonyan, K. & Zisserman, A. Very deep convolutional networks for large-scale image recognition. *3rd Int. Conf. Learn. Represent. ICLR 2015 - Conf. Track Proc.* 1–14 (2015).

43. Fialka, O. & Čadík, M. FFT and convolution performance in image filtering on GPU. *Proc. Int. Conf. Inf. Vis.* 609–614 (2006) doi:10.1109/IV.2006.53.

44. Krizhevsky, A., Sutskever, I. & Hinton, G. E. ImageNet classification with deep convolutional neural networks. *Commun. ACM* (2017) doi:10.1145/3065386.

45. Szegedy, C. *et al.* Going deeper with convolutions. *Proc. IEEE Comput. Soc. Conf. Comput. Vis. Pattern Recognit.* **07-12-June**, 1–9 (2015).

46. Al-Ashrafy, M., Salem, A. & Anis, W. An efficient implementation of floating point multiplier. *Saudi Int. Electron. Commun. Photonics Conf. 2011, SIECPC 2011* (2011) doi:10.1109/SIECPC.2011.5876905.

47. Gao, L., Chen, P. Y. & Yu, S. Demonstration of Convolution Kernel Operation on Resistive Cross-Point Array. *IEEE Electron Device Lett.* **37**, 870–873 (2016).

48. Shafiee, A. *et al.* ISAAC: A Convolutional Neural Network Accelerator with In-Situ Analog Arithmetic in Crossbars. *Proc. - 2016 43rd Int. Symp. Comput. Archit. ISCA 2016* 14–26 (2016) doi:10.1109/ISCA.2016.12.

49. Li, X. *et al.* Fast and reliable storage using a 5 bit, nonvolatile photonic memory cell. *Optica* **6**, 1 (2019).

50. Ríos, C. *et al.* Integrated all-photonic non-volatile multi-level memory. *Nat. Photonics* **9**, 725–732 (2015).

51. Feldmann, J. *et al.* Calculating with light using a chip-scale all-optical abacus. *Nat. Commun.* **8**, (2017).

52. Ríos, C. *et al.* In-memory computing on a photonic platform. *Sci. Adv.* **5**, (2019).

53. Gehring, H. *et al.* Low-loss fiber-to-chip couplers with ultrawide optical bandwidth. *APL Photonics* **4**, 0–7 (2019).

54. Gehring, H., Eich, A., Schuck, C. & Pernice, W. H. P. Broadband out-of-plane coupling at visible wavelengths. *Opt. Lett.* **44**, 5089 (2019).

55. Gaeta, A. L., Lipson, M. & Kippenberg, T. J. Photonic-chip-based frequency combs. *Nature Photonics* vol. 13 158–169 (2019).

56. Pfeiffer, M. H. P. *et al.* Photonic Damascene Process for Integrated High-Q Microresonator Based Nonlinear Photonics. *Optica* **3**, 1–6 (2016).

57. Grother, P. J. & Hanaoka, K. K. NIST Special Database 19 - Handprinted Forms and Characters Database. *Tech. Rep. Spec. Database 19* 1–30 (2016) doi:10.18434/T4H01C.

58. Ma, Y. *et al.* Ultralow loss single layer submicron silicon waveguide crossing for SOI optical interconnect. *Opt. Express* **21**, 29374 (2013).

59. Lu, Z. *et al.* Broadband silicon photonic directional coupler using asymmetric-waveguide based phase control. *Opt. Express* **23**, 3795 (2015).

60. Nahmias, M. A. *et al.* Photonic Multiply-Accumulate Operations for Neural Networks.

*IEEE J. Sel. Top. Quantum Electron.* (2019) doi:10.1109/jstqe.2019.2941485.

61. Farmakidis, N. *et al.* Plasmonic nanogap enhanced phase change devices with dual electrical-optical functionality. *Sci. Adv.* **5**, 1–8 (2019).

62. Zhang, H. *et al.* Miniature Multilevel Optical Memristive Switch Using Phase Change Material. *ACS Photonics* **6**, 2205–2212 (2019).

63. Atabaki, A. H. *et al.* Integrating photonics with silicon nanoelectronics for the next generation of systems on a chip. *Nature* **556**, (2018).

64. Wang, X. & Liu, J. Emerging technologies in Si active photonics. *J. Semicond.* **39**, (2018).

65. Sun, J., Timurdogan, E., Yaacobi, A., Hosseini, E. S. & Watts, M. R. Large-scale nanophotonic phased array. *Nature* **493**, 195–199 (2013).

66. Gehring, H., Blaicher, M., Hartmann, W. & Pernice, W. H. P. Python based open source design framework for integrated nanophotonic and superconducting circuitry with 2D-3D-hybrid integration. *OSA Contin.* **2**, 3091–3101 (2019).

67. Liu, J. *et al.* Ultralow-Power Chip-Based Soliton Microcombs for Photonic Integration. *Optica* **5**, (2019).

68. Guo, H. *et al.* Universal dynamics and deterministic switching of dissipative Kerr solitons in optical microresonators. *Nat. Phys.* (2017) doi:10.1038/nphys3893.

69. Karpov, M. *et al.* Dynamics of soliton crystals in optical microresonators. *Nat. Phys.* (2019) doi:10.1038/s41567-019-0635-0.

# Methods

**Device fabrication**

The photonic circuits used for the convolution experiments are fabricated using a three-step electron-beam lithography (EBL; Raith EBPG 5150) process on a silicon nitride (325 nm) on silicon oxide (3300 nm) on silicon wafer (Rogue Valley Microdevices). The complete circuit was designed using GDShelpers, a design framework for integrated circuitry[66].

In the first lithography step, windows in the positive tone resist Polymethylmethacrylat (PMMA) are exposed for the deposition of alignment markers made from gold. The resist is developed in 1:3 MIBK:Isopropanol for 120 seconds and a layer stack of 5 nm chromium, 120 nm gold and 5 nm chromium are evaporated via electron-beam physical vapour deposition (EBPVD). By sonicating the chip in acetone, the PMMA is removed and only the gold markers in the exposed positions remain. The markers are used in the second step to align the photonic structures. After spin coating a layer of 300 nm of the resist and prebaking it for 60 seconds at 85°C, an etch mask is exposed in the negative-tone ebeam resist arN 7520.12. The photonic structures are developed in MF-319 for 75 seconds and a post-development bake is performed at 85°C for 60 seconds. Using reactive ion etching with a CHF3/O2 plasma the mask of the photonic circuits is transferred into the sample. The silicon nitride layer is fully etched leaving single mode waveguides at telecom-wavelengths with a width of 1.2 µm and a height of 325 nm. Subsequently the remaining resist is removed in an oxygen plasma for 10 minutes. In the third EBL step, windows for the deposition of the phase-change material are written using the same markers as for the photonic structures for the alignment. The same process as in the first EBL step is used. Finally, 10 nm of the phase-change material GST and 10 nm of indium tin oxide (ITO) are sputter deposited on the sample. Both layers are sputtered using RF sputtering with an argon plasma (5 mtorr pressure, 15 sccm Ar, 30 W RF power and a base pressure of $2\times10^{-6}$ Torr). The ITO is used as a protective film to prevent oxidation of the phase-change material. As in the marker-deposition, the PMMA is lifted off by sonicating the sample in acetone leaving the phase-change material only in the desired positions on the photonic circuitry. Prior to the experiments the GST is crystallized on a hot plate at 220°C for approximately 10 minutes.

**Measurement setup**

The experimental setups used to perform the convolution experiments are shown in supplementary figures S1-S3. The individual wavelengths are generated using a frequency comb that is operated in the single soliton state and separated using a fibre-based multiplexer. For the image processing experiments (Fig. 3 and 4) the wavelengths (input vectors) are modulated using variable optical attenuators based on micro-electro-mechanical systems (MEMS), whereas the fast modulation (Fig. 5) was performed with a 20 GHz electro-optic modulator (EOM). The input signal is coupled to the chip using 3D printed broadband total internal reflexion couplers capable of operating from the visible to the telecom wavelengths regime.

In the multiplexed version of the experiment processing four vectors at the same time, the corresponding wavelengths are multiplexed and demultiplexed accordingly before and after the matrix again using fiber-multiplexers. The convolution results are read using photodetectors (New Focus Model 2011). In the frequency response experiment (Fig. 5) a fast photodiode (12 GHz) was utilised.

The measurement setup remains stable for extended periods, as also detailed in the supplementary information. Fluctuations in the transmission are due to temperature variations in the laboratory, which oscillate during day and night times. The long-term trend, however, remains unchanged over weeks (see S7) and also months[50].

**Realization of high-Q Si3N4 microresonators**

The soliton microcombs used in our work are based on $Si_3N_4$ microring resonators with free spectral range (FSR) of 100 GHz shown in Fig. 2 b. The microresonators are fabricated using Photonic Damascene process [56], which provides access to high Q factors reaching $10^7$ and enables the four-wave-mixing based nonlinear frequency conversion processes as well as the formation of DKS states at low pump powers [67].

The microresonators were designed to have cross-section dimensions of 0.82 x 1.50 µm, which ensure anomalous group velocity dispersion (GVD) of about 1-2 MHz at around 1550 nm needed for the Kerr comb generation and the formation of DKS states. The light is coupled evanescently to a microresonator via the on-chip bus waveguide with similar dimensions located close to the microring, and which are additionally equipped with inverse tapers at the ends for edge chip coupling. Employed $Si_3N_4$ chips are furthermore fiber-packaged with average loss 4 dB/interface to facilitate the light coupling in and out of the system. The fabricated devices have Q-factors exceeding $5 \times 10^6$, which allows for the DKS generation and switching [68,69] even for relatively low input pump powers below 1W.

**Soliton comb generation**

For the DKS generation a $Si_3N_4$ microring resonator is driven using continuous wave tunable fiber laser which is amplified with an Erbium-doped fiber amplifier (EDFA) to the power level of about 1 W. A high-power bandpass filter is used to suppress the amplified spontaneous emission (ASE) from the EDFA. The light polarization is adjusted using fiber-based polarization controller to match the TE-polarized fundamental mode of the microresonator, and then is launched to the fiber-coupled $Si_3N_4$ chip.

In order to launch the DKS state, a standard pump tuning technique is applied [32] where the amplified seed laser is swept over the choosen frequency resonance from the blue-detuned side to the red detuned side at a speed of approximately 200 GHz/s. This approach allows to generate multiple-soliton states with several pulses inside the cavity, which however usually has highly structures optical spectrum. In order to achieve the single DKS state with spectrally smooth sech$^2$–shape envelope the soliton switching procedure is employed [68] and pump is slowly tuned toward shorter wavelength until the single soliton state is stabilized. In order to improve the long-term stability of the generated DKS states and align the resulting optical frequency comb to the established International Telecommunication Union (ITU) grids, the $Si_3N_4$ chip is thermally controlled which enables the usage of the standard WDM equipment and optical comb stabilization against environmental temperature fluctuations and setup drifts ensuring > 8 hours of continuous operation.

The resulting DKS-based optical frequency comb with 100 GHz line spacing and spanning over multiple telecommunication bands is coupled out from the chip. The residual pump is suppressed using fiber-based notch filter, and a small portion of the light (1 %) is used for the monitoring purposes. The rest of the comb is shown in Fig.2 e, and is then additionally amplified with C-band EDFA to further employ it in the setup for image vectors encoding and demultiplexing. The amplification of the EDFA of up to 15 dB was individually chosen for the different experiments and is mainly used to compensate for coupling losses between the fibre array and the chip.

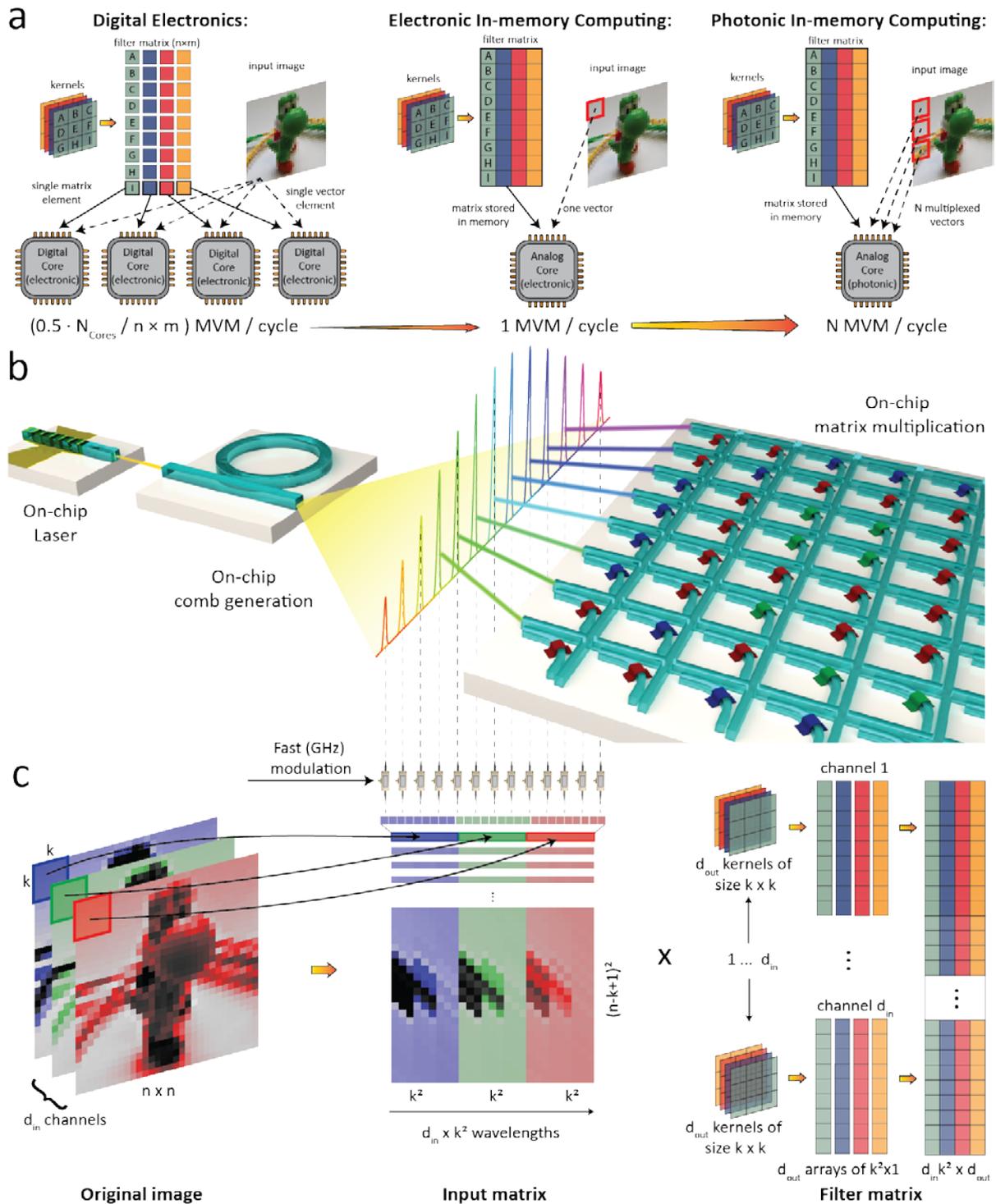

**Figure 1. Photonic in-memory computing using an on-chip frequency comb and phase-change materials. a)** A comparison of digital and analog electronic architectures with our photonic tensor core architecture. Digital electronics (left) requires many sequential processing steps distributed across multiple cores to compute convolution operations on an image, while an entire matrix-vector multiplication (MVM) can be performed in one step using analog

electronic in-memory computing (center). Photonic in-memory computing brings wavelength multiplexing as an additional degree of freedom, enabling multiple MVM operations in a single time step. **b)** Conceptual illustration of a fully integrated photonic architecture to compute convolution operations. An on-chip laser (not used) pumps an integrated SiN soliton microcomb to generate a broadband frequency comb. Individual comb teeth which form the input vectors are modulated at high speeds, multiplied with a matrix of non-volatile phase-change memory cells, and summed along each column on a photodetector. **c)** An input image with $d_{in}$ channels is convolved with $d_{out}$ kernels of size $k \times k$ by mapping convolution operations into a sequence of MVM operations. The input image is mapped to a series of $(n-k+1)^2$ input vectors of size $(d_{in} \times k^2) \times 1$ and multiplied by a filter matrix of dimension $(d_{in} \times k^2) \times d_{out}$. Each comb line corresponds to one entry of the input vector and is modulated according to the pixel values of the input matrix.

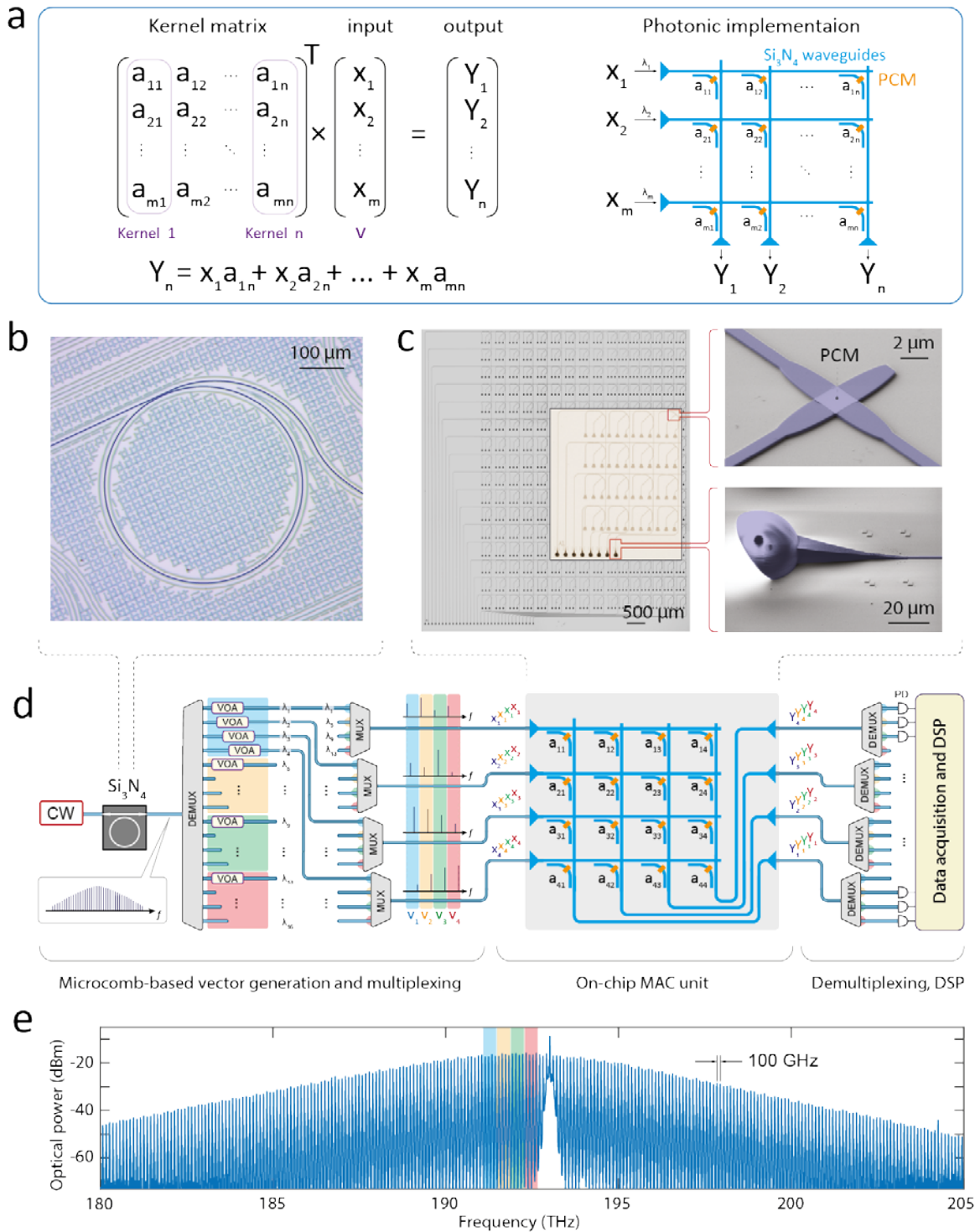

**Figure 2. Concept of photonic tensor cores for convolution operations. a)** Basic matrix-vector multiplication: A vector is encoded in the amplitude of individual comb teeth of a silicon nitride ($Si_3N_4$) photonic integrated soliton frequency comb (microcomb) exhibiting wavelengths ("$X_1$" to "$X_m$") and send to the corresponding matrix input waveguides. The matrix elements are inscribed in the state of phase-change material patches on the waveguides. The splitting ratios of the directional couplers are chosen such that the same fraction of the light for each input reaches the output. **b)** Scanning electron miccograph of a microresonator used for frequency comb generation **c)** Optical micrograph of a fabricated 16×16. The inset shows a 4x4 matrix with 3D printed input and output couplers to enable broadband operation. The close-up SEM images on the right show the 3D printed couplers (bottom) and the waveguide crossings with the PCM (top) in more detail. **d)** Sketch of the multiplexed all-optical matrix-vector multiplication. The input vectors are generated from lines of a photonic chip-scale dissipative Kerr soliton (DKS) frequency comb using a multiplexer (MUX) and variable optical attenuators (VOAs). The entries of different input vectors are grouped together again employing wavelength multiplexing and sent to the on-chip MAC-unit (Multiply-Accumulate-unit) that performs the calculations. After combining the correct wavelengths with optical demultiplexers (DEMUX), the multiplication results are be obtained. Note that in the given example four kernels and four input vectors are operated at once, resulting in 64 MAC-operations per time step. **e)** Measured spectrum of a single-soliton frequency comb.

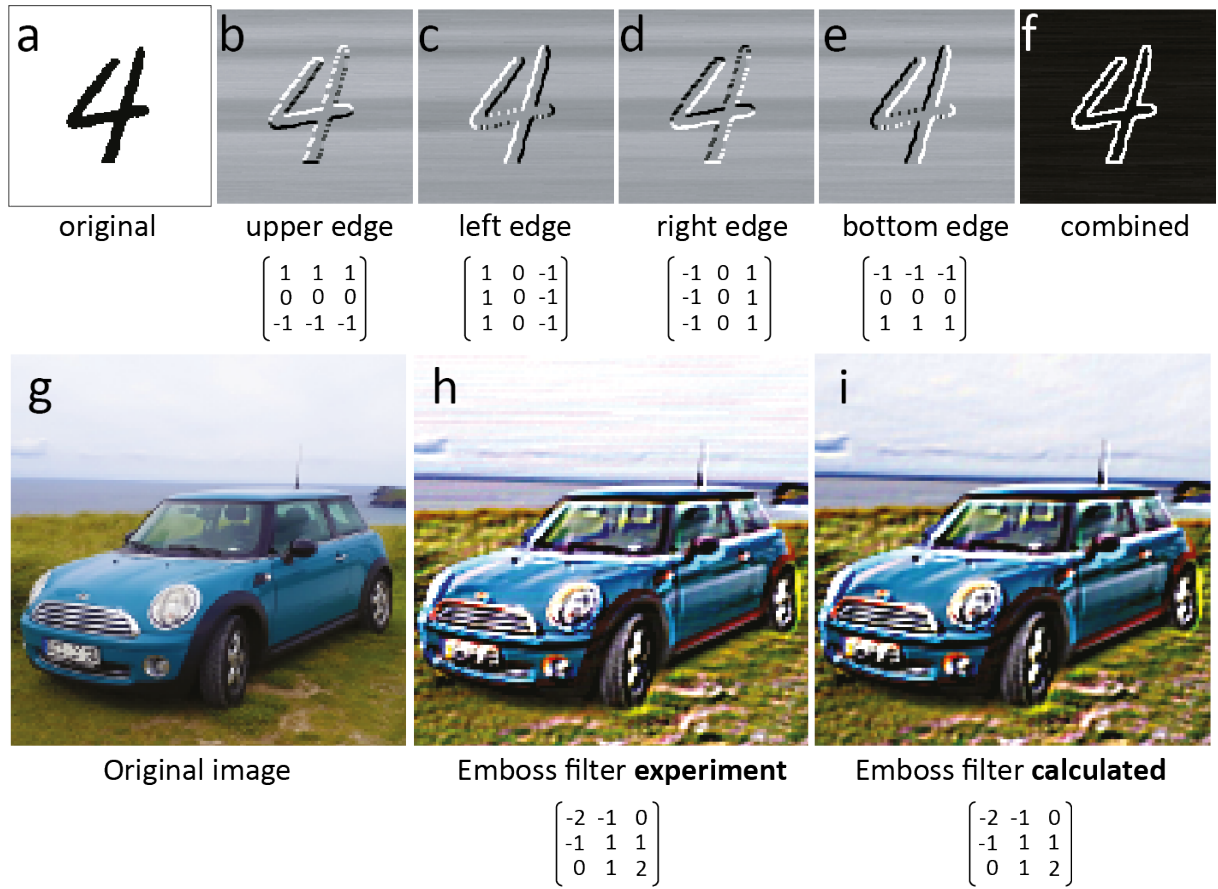

**Figure 3. Convolution using sequential MVM operations. a-e)** Experimental result of convolving a 128×128 pixel image showing a handwritten digit (a) with four image kernels of the size 3×3 (corresponding to a 9×4 filter matrix). The kernels are chosen to highlight different edges of the input image. **f)** Combined image from b-e showing edge highlighting. **g)** Convolution operation with a 3×3 sized image kernel (i.e. emboss filter) without post-processing. The image on the left shows the original image while the other two depict the experimental and the calculated (correct) result.

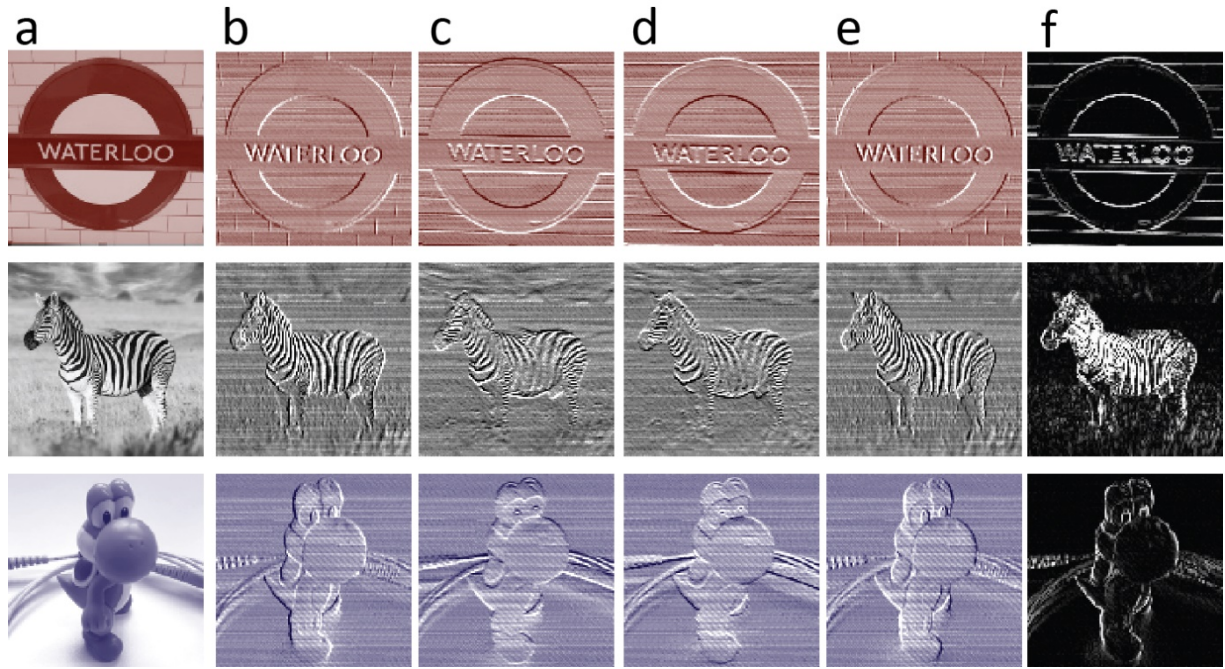

**Figure 4. Convolution using parallel MVM operations. a-e)** The original input images are shown on the left (a) and the output images using four different image kernels for highlighting edges are shown in (b)-(e). The size of the four image kernels is 2×2 corresponding to a 4×4 filter matrix. In each time step, four input vectors are processed simultaneously via wavelength division multiplexing as illustrated in Fig. 2c. **f)** Combined image from b-e showing successful edge highlighting.

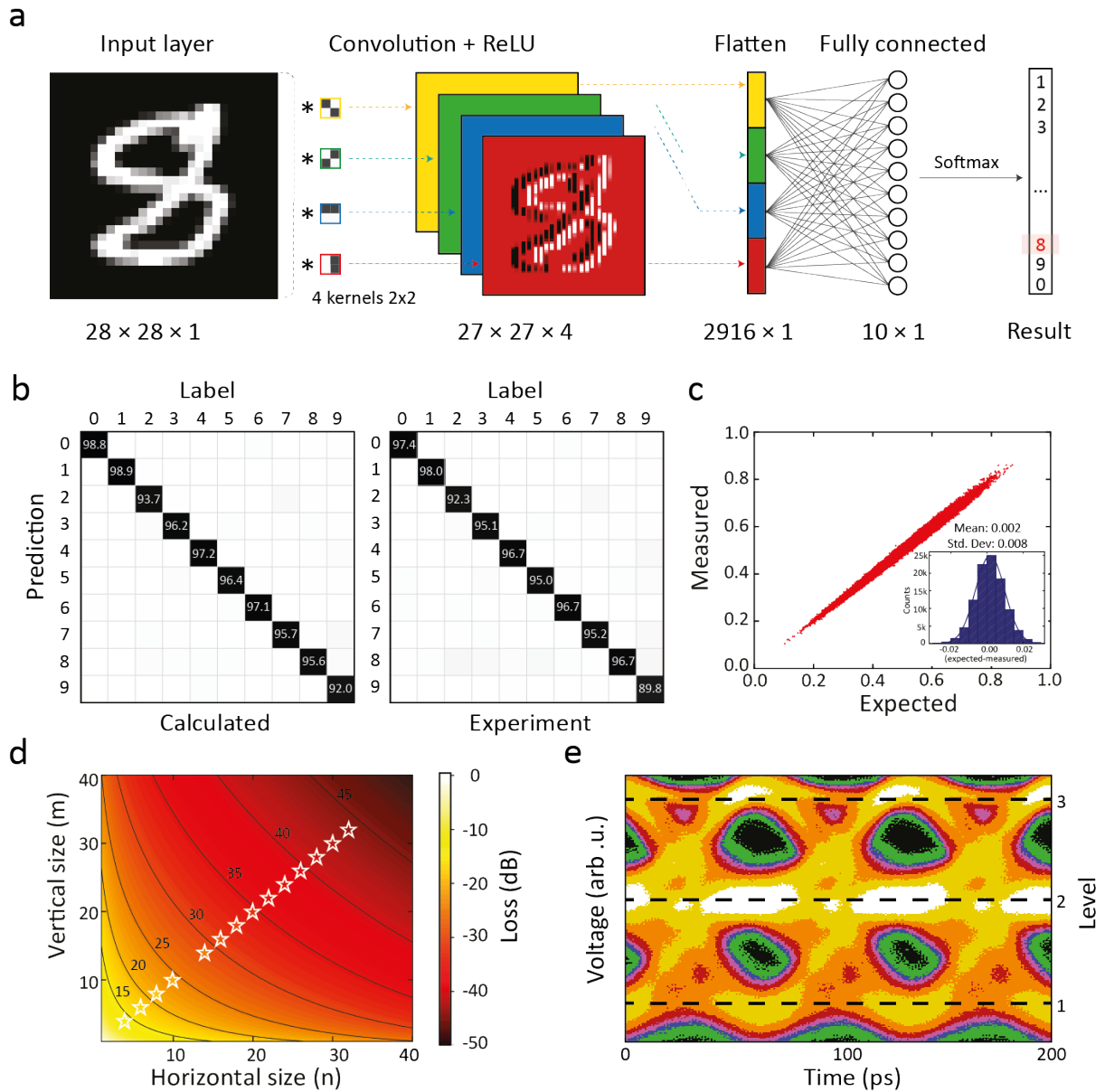

**Figure 5. Digit recognition with a convolutional neural network and scalability. a)** Layer structure of the network used to test the photonic tensor core with the MNIST handwritten digits database **b)** Confusion matrices showing similar performance for the prediction results for the experimental (95.3%) and calculated CNN (96.1%). **c)** Calculation accuracy for 100,000 MAC operations multiplying a vector of nine entries with a fixed matrix. Inset: Histogram of the data revealing a standard deviation of 0.008 and therefore a resolution of 5 bit. **d)** Optical loss of the matrix as a function of its size. The heatmap depicts calculated optical loss for a directional coupler loss of 0.1 dB and a crossing loss of 0.12 dB. The stars are measured optical loss for fabricated matrices. **e)** Eye-diagram for a matrix multiplication with a 2x1 matrix at 13.5 GHz

modulation speed. The two inputs are modulated with two PRBS patterns resulting in three different levels for the multiplication result.